\documentclass[aps,twocolumn,showpacs]{revtex4}
\begin{document}
\title{Towards a relativistic statistical theory}
\author{G. Kaniadakis}
\email{giorgio.kaniadakis@polito.it} \affiliation{Dipartimento di
Fisica, Politecnico di Torino, \\ Corso Duca degli Abruzzi 24,
10129 Torino, Italy}
\date{\today}

\begin{abstract}
In special relativity the mathematical expressions, defining
physical observables as the momentum, the energy etc,  emerge as
one parameter (light speed) continuous deformations of the
corresponding ones of the classical physics. Here, we show that
the special relativity imposes a proper one parameter continuous
deformation also to the expression of the classical
Boltzmann-Gibbs-Shannon entropy. The obtained relativistic
entropy permits to construct a coherent and selfconsistent
relativistic statistical theory [Phys. Rev. E {\bf 66}, 056125
(2002);  Phys. Rev. E {\bf 72}, 036108 (2005)], preserving the
main features (maximum entropy principle, thermodynamic stability,
Lesche stability, continuity, symmetry, expansivity, decisivity,
etc.) of the classical statistical theory, which is recovered in
the classical limit. The predicted distribution function is a
one-parameter continuous deformation of the classical
Maxwell-Boltzmann distribution and has a simple analytic form,
showing power law tails in accordance with the experimental
evidence.

\end{abstract}
 \maketitle

\section{Introduction}

In the ordinary relativistic statistical mechanics the entropy is
assumed to have the same form  of the entropy of classical
statistical mechanics. Then, starting from the
Boltzmann-Gibbs-Shannon entropy, the maximum entropy principle
yields an exponential distribution which exactly reproduces the
Maxwell-Boltzmann distribution of classical statistical mechanics
in the rest frame. Unfortunately the Maxwell-Boltzman
distribution fails to explain the spectrum of cosmic rays, which
undoubtedly represent the most famous and important relativistic
particle system \cite{Vasyliunas,Biermann}. This spectrum has a
very large extension (13 decades in energy and 33 decades in
particle flux), exhibiting a power law asymptotic behavior.
Similar power law asymptotic behaviors of the distribution
function have been observed also in other relativistic systems
including plasmas \cite{Hasegawa} and multiparticle production
processes \cite{Wilk}. Hence for relativistic systems, the
experimental evidence suggests a non-exponential distribution
function with power law tails.  In the last 40 years, empirical
non-exponential distributions with power law tails  have been
systematically used in high energy plasmas.

A non-exponential distribution can be originated exclusively by
an entropy manifestly different from the Boltzmann-Gibbs-Shannon
one. In order to propose theories based on non-exponential
statistical distributions, one can follow different paths, that
allow to construct, other coherent and selfconsistent statistical
theories besides the classical statistical mechanics
\cite{Gell-Mann,Abe,Naudts,Chavanis,Bashkirov,PhA01,PRE02,PRE05}.

We recall that any relativistic formula, defining a physical
quantity, e.g. momentum, energy, etc, can be viewed as a one
parameter (light speed) generalization or deformation of the
corresponding classical formula. Consequently, it is natural to
ask if the entropy of a relativistic system could be a one
parameter generalization of the Boltzmann-Gibbs-Shannon entropy
as well.

The main goal of the present contribution is to show that the
one-particle relativistic dynamics imposes a one parameter
generalization of the Boltzmann-Gibbs-Shannon entropy. The new
entropy permits the construction of a statistical mechanics which
efficiently describes  the relativistic many-particle systems.

\section{ Composition laws in special relativity}

We introduce the dimensionless variables momentum $q$, velocity
$u$ and total energy ${\cal E}$ in place of the corresponding
physical variables $p$, $v$ and $E$ through
\begin{eqnarray}
\frac{v}{u}=\frac{p}{m q}=\sqrt{\frac{E}{m{\cal E}}}=\kappa c=v_*
\ \label{RII16} \ \ ,
\end{eqnarray}
being $c$ the light speed and $\kappa$ a dimensionless parameter.
Furthermore we impose for the velocity $v_*$ that
$\lim_{c\rightarrow \infty,\,\, \kappa\rightarrow 0}
v_*=v_*^{\infty}$ with $v_*^{\infty}=$ finite, in order to
preserve the validity of the dimensionless variable definitions,
also in the classical limit. Within the special relativity it is
immediate to express the deformation effects, due to the finite
value of the light speed, in terms of the dimensionless parameter
$\kappa$:
\begin{eqnarray}
&&q(u)=\frac{u}{\sqrt{1-\kappa^2u^2}} , \label{RII7} \\
&&u(q)=\frac{q}{\sqrt{1+\kappa^2q ^2}} \ \ , \label{RII2}  \\
&&{\cal E}(q)=\frac{1}{\kappa^2}\sqrt{1+\kappa^2q^2} ,
\label{RII14}
\\ &&q({\cal E})=\sqrt{\kappa^2{\cal E}^2-\frac{1}{\kappa^2}} , \label{RII9} \\
&&{\cal E}(u)=\frac{1}{\kappa^2\sqrt{1-\kappa^2u^2}} ,
\label{RII15}\\
&&u({\cal E})=\frac{\sqrt{\kappa^2{\cal
E}^2-1/\kappa^2}}{\kappa^2{\cal E}} \ \ . \label{RII11}
\end{eqnarray}

Consider in the one-dimensional frame $\cal S$ two identical
particles of rest mass $m$. We suppose that the first particle
moves towards right with velocity $u_1$ while the second particle
moves towards left with velocity $u_2$. The momenta of the two
particles are given by $q_1=q\,(u_1)$ and $q_2=q\,(u_2)$
respectively, where $q\,(u)$ is defined through Eq. (\ref{RII7}).
The energies of the two particles are given by ${\cal E}_1={\cal
E}\,(u_1)$ and ${\cal E}_2={\cal E}\,(u_2)$ respectively, with
${\cal E}\,(u)$ defined by Eq. (\ref{RII15}). We consider now the
same particles in a new frame $\cal S\,'$ which moves at constant
speed $u_2$ towards left with respect to the frame $\cal S$. In
this new frame, which is the rest frame of the second particle,
we have that the two particles have velocities given by
$u\,'_1=u_1\stackrel{\kappa}{\oplus}u_2$ and $u\,'_2=0$
respectively, being
\begin{equation}
u_1\stackrel{\kappa}{\oplus}u_2=\frac{u_1+u_2}{1+\kappa^2u_1u_2}
\label{RII19} \ \ ,
\end{equation}
the well known relativistic additivity law for the dimensionless
velocities. In the same frame $\cal S\,'$ the particle momenta
are given by $q\,'_1=q\,(u\,'_1)$ and $q\,'_2=0$, respectively.
Analogously in $\cal S\,'$ the particle energies of the two
particle in $\cal S\,'$ results ${\cal E}\,'_1={\cal
E}\,(u\,'_1)$ and ${\cal E}\,'_2=1/\kappa^2$. A very interesting
result follows from the relation ${\cal E}\,'_2=1/\kappa^2$
regarding the physical meaning of the deformation parameter
$\kappa$. It is evident that the quantity $1/\kappa^2$ represents
the dimensionless rest energy of the relativistic particle.

Let us pose now the following questions: is it possible to obtain
the value of the momentum $q\,'_1$ (of the energy ${\cal
E}\,'_1$) starting directly from the values of the momenta $q_1$
and $q_2$ (of the energies ${\cal E}\,_1$ and ${\cal E}\,_2$) in
the frame $\cal S$. The answers to the above questions are
affirmative.

First we consider the case of the relativistic momentum $q\,'_1$
and write it as
$q\,'_1=q\,(u\,'_1)=q(u_1\!\!\stackrel{\kappa}{\oplus}\!u_2)$. In
ref. \cite{PRE02} it has been shown that $q\,'_1
=q_1\stackrel{\kappa}{\oplus}q_2$ being
\begin{eqnarray}
q_1\stackrel{\kappa}{\oplus}q_2 = q_1\sqrt{1\!+\!\kappa^2q_2^2}
+q_2\sqrt{1\!+\!\kappa^2q_1^2} \ \ , \label{RII22}
\end{eqnarray}
the $\kappa$-sum of relativistic momenta. In words, the
relativistic momentum $q\,'_1$ of the first particle, in the rest
frame of the second particle $\cal S'$, is the $\kappa$-deformed
sum of the momenta $q_1$ and $q_2$ of the two particles in the
frame $\cal S$. The $\kappa$-sum of the relativistic momenta and
the relativistic sum of the velocities are intimately related,
and reduce both to the standard sum as the velocity $c$
approaches to infinity (or equivalently the parameter $\kappa$
approaches to zero). The deformations in the above sums, in both
cases, are relativistic effects and are originated from the fact
that $c$ has a finite value.

We consider now the total energy ${\cal E}\,'_1$ of the first
particle in the frame $\cal S\,'$.  Clearly it results ${\cal
E}\,'_1={\cal E}(u_1')={\cal E}\left({ u
}_1\stackrel{\kappa}{\oplus}{u }_2\right)$. In ref. \cite{PRE05}
it is shown that one can calculate ${\cal E}\,'_1$ starting
directly from the values, in the frame $\cal S$, of the total
energies ${\cal E}\,_1$ and ${\cal E}\,_2$. It is obtained ${\cal
E}\,'_1= {\cal E}_1\stackrel{\kappa}{\oplus}{\cal E }_2$ where
the composition law of the dimensionless relativistic total
energies is defined through
\begin{eqnarray}
{\cal E}_1\stackrel{\kappa}{\oplus}{\cal E}_2 = \kappa^2{\cal
E}_1{\cal E}_2\!+\!\frac{1}{\kappa^2}\sqrt{\big(\kappa^4{\cal
E}_1^2\!-\!1\big)\big(\kappa^4{\cal E}_2^2\!- \!1\big)}\ \ .
\label{RII26}
\end{eqnarray}

In the following we summarize the relationships linking the
$\kappa$-sums defined through Eqs. (\ref{RII19}), (\ref{RII22})
and (\ref{RII26})
\begin{eqnarray}
&&{q }(u_1)\stackrel{\kappa}{\oplus}{q }(u_2) ={q }\left({ u
}_1\stackrel{\kappa}{\oplus}{u }_2\right) \ \ , \label{RII32} \\
&&{\cal E }(u_1)\stackrel{\kappa}{\oplus}{\cal E}(u_2) ={\cal E
}\left({u}_1\stackrel{\kappa}{\oplus}{u}_2\right) \ \ , \label{RII33} \\
&&{u }(q_1)\stackrel{\kappa}{\oplus}{u}(q_2) ={u
}\left({q}_1\stackrel{\kappa}{\oplus}{q}_2\right) \ \ , \label{RII35} \\
&&{\cal E }(q_1)\stackrel{\kappa}{\oplus}{\cal E}(q_2) ={\cal E
}\left({q}_1\stackrel{\kappa}{\oplus}{q}_2\right) \ \ , \label{RII36} \\
&&q({\cal E}_1)\stackrel{\kappa}{\oplus}q({\cal E}_2) ={q
}\left({\cal E}_1\stackrel{\kappa}{\oplus}{\cal E}_2\right) \ \ , \label{RII38} \\
&&u({\cal E}_1)\stackrel{\kappa}{\oplus}u({\cal E}_2) ={u
}\left({\cal E}_1\stackrel{\kappa}{\oplus}{\cal E}_2\right) \ \ .
\label{RII39}
\end{eqnarray}
Note that in the lhs and rhs of any of the above equations appear
two different composition laws. For instance in Eq.(\ref{RII32})
in the lhs appears the composition law of the dimensionless
momenta (\ref{RII22}) while in the rhs appears the composition
law of the dimensionless velocities (\ref{RII19}). We emphasize
that the three $\kappa$-sums given by Eqs. (\ref{RII19}),
(\ref{RII22}) and (\ref{RII26}), emerge only when we change the
frame of observation of the particle from $\cal S$ to $\cal S'$.

\section{Lorentz invariant integration}

Within the special relativity a central role is played by the
four dimension Lorentz invariant integral
\begin{eqnarray}
I_{rel}=A \int d^4p \,\,
\theta(p_0)\,\delta(p^{\mu}p_{\mu}-m^2c^2) \, \,F \ \
,\label{RIV4}
\end{eqnarray}
with $A$ an arbitrary constant and $F$ an arbitrary function
depending on $p=|p|$. After recalling that
$p^{\mu}=(E/c,\,\mbox{\boldmath $p$}\,)$ and $E=\sqrt{m^2
c^4+p^{2} c^2}$, one can reduce the above integral as follows
\begin{eqnarray}
I_{rel}\!=\!\int \!\frac{d^3 q}{\sqrt{1+\kappa^2 q^2 }}\,F \!=\!
\int_0^{\infty}\!\! \frac{d q}{\sqrt{1+\kappa^2 q^2 }}\,4\pi\, q^2
\,F  \ \ , \label{RIV5}
\end{eqnarray}
where the constant $A$ is fixed properly and the dimensionless
momentum is introduced. We remark that in (\ref{RIV4}) the
integration element $d^4p$ is a scalar because the Jacobian of
the Lorentz transformation is equal to unity, so that $I_{rel}$
transforms as $F$. Then in (\ref{RIV5}) the integration element
$d^3 q/\sqrt{1+\kappa^2 q^2 }$ is a scalar. After introducing the
$\kappa$-differential through
\begin{eqnarray}
d_{{\scriptscriptstyle\{}\kappa{\scriptscriptstyle\}}}q=\frac{d
q}{\sqrt{1+\kappa^2 q^2 }} \ \ . \label{RV8}
\end{eqnarray}
the integral $I_{rel}$ assumes the form
\begin{eqnarray}
I_{rel}=\int_0^{\infty}d_{{\scriptscriptstyle\{}\kappa{\scriptscriptstyle\}}}q
\,\,4\pi\, q^2 \,F \ \ . \label{RIV8}
\end{eqnarray}
One immediately observes that the Lorentz invariant integral can
be obtained by deforming the ordinary (classical) integral
\begin{eqnarray}
I_{cl}=\int_0^{\infty}dq \,\,4\pi\, q^2 \,F \ \ . \label{RIV8}
\end{eqnarray}
The deformation is obtained by performing the substitution $dq
\rightarrow
d_{{\scriptscriptstyle\{}\kappa{\scriptscriptstyle\}}}q$.

It is important to note that the $\kappa$-differential (\ref{RV8})
can be obtained  directly from the additivity law of the
dimensionless relativistic momenta (\ref{RII22}) according to
\begin{eqnarray}
d_{{\scriptscriptstyle\{}\kappa{\scriptscriptstyle\}}}q=
(q+dq)\stackrel{\kappa}{\ominus}q \ \ . \label{RII22a1}
\end{eqnarray}

\section{Physical meaning of the $\kappa$-differential}

We consider now two identical relativistic particles with rest
mass $m$ and velocities ${\bf v}_1$ and ${\bf v}_2$ respectively,
in the frame ${\cal S}$. The modulus of the relative velocity
$V=V({\bf v}_1,{\bf v}_2)$ of the two particles, given in ref.
\cite{Cercignani} (page 20), can be written also in the form
\begin{eqnarray}
V({\bf v}_1,{\bf v}_2)= \sqrt{\left({\bf v}_1\ominus {\bf
v}_2\right)^2-\frac{1}{c^2}\left(\frac{{\bf v}_1 \times{\bf
v}_2}{1-{\bf v}_1{\bf v}_2/c^2}\right)^2} , \label{RV1}
\end{eqnarray}
with
\begin{eqnarray}
{\bf v}_1\ominus {\bf v}_2= \frac{{\bf v}_1 - {\bf v}_2}{1-{\bf
v}_1{\bf v}_2/c^2} \ \ . \label{RV11}
\end{eqnarray}

We perform now the calculation of the modulus of relative
momentum  of the two particles according to $P(V)=m
V/\sqrt{1-V^2/c^2}$. Clearly, it results that $P(V)=P({\bf
v}_1,{\bf v}_2)$ and then
 $P(V)=P({\bf p}_1,{\bf p}_1)$. At this point
we introduce the dimensionless variables ${\bf q}={\bf p}/mv_{*}$
and $Q=P/mv_{*}$. After some straightforward calculation, one
arrives to the following expression for the modulus of the
dimensionless relative momentum (the modulus of dimensionless
momentum of the particle 1 (2) in the rest frame of other
particle)
\begin{eqnarray}
{\rm Q}({\bf q}_1,{\bf q}_2)= \sqrt{\frac{\left({\bf q}_1
\stackrel{\kappa}{\ominus}{\bf q}_2\right)^2-\kappa^2\left({\bf
q}_1 \times{\bf q}_2\right)^2}{1+\kappa^4\left({\bf q}_1
\times{\bf q}_2\right)^2} } \ \ , \label{RV5}
\end{eqnarray}
with
\begin{eqnarray}
{\bf q}_1\stackrel{\kappa}{\ominus}{\bf q}_2 = {\bf
q}_1\sqrt{1\!+\!\kappa^2{\bf q}_2^2} -{\bf
q}_2\sqrt{1\!+\!\kappa^2{\bf q}_1^2} \ \ . \label{RV51}
\end{eqnarray}
The expression of the relative momentum given by Eq. (\ref{RV5})
has been obtained in \cite{PRE05} (there the formula contains a
typing error).

In the following, in order to explain the physical meaning of the
$\kappa$-differential and the $\kappa$-derivative in special
relativity,  we consider that the two particles move along the
same line so that ${\bf q}_1 \times{\bf q}_2=0$.

We suppose now that the two particles have momenta ${\bf
q}_1={\bf q}+d{\bf q}$ and ${\bf q}_2={\bf q}$ respectively in
the frame ${\cal S}$. To calculate the modulus of the momentum
{\rm Q}({\bf q}+d{\bf q},\,{\bf q}) of one particle in the rest
frame of the other particle, after posing $q=|{\bf q}|$, one
obtains:
\begin{eqnarray}
{\rm Q}({\bf q}+d{\bf q},{\bf
q})=d_{{\scriptscriptstyle\{}\kappa{\scriptscriptstyle\}}}q  \ \
. \label{RV8a}
\end{eqnarray}
The physical meaning of the $\kappa$-differential
$d_{{\scriptscriptstyle\{}\kappa{\scriptscriptstyle\}}}q$
immediately follows.  The infinitesimal difference $dq$ of the
momenta of the particle $1$ and $2$ in the frame ${\cal S}$
becomes $d_{{\scriptscriptstyle\{}\kappa{\scriptscriptstyle\}}}q$
if this difference is observed in the rest frame of one of the two
particles. The meaning of the $\kappa$-derivative
\begin{eqnarray}
\frac{d}{d_{{\scriptscriptstyle\{}\kappa{\scriptscriptstyle\}}}q}
=\sqrt{1+\kappa^2 q^2}\, \frac{d}{dq} \ \ , \label{RV8b}
\end{eqnarray}
readily follows. If $d/dq$ represents the derivative in the frame
${\cal S}$, the deformed derivative
$d/d_{{\scriptscriptstyle\{}\kappa{\scriptscriptstyle\}}}q$
represents an ordinary derivative in the rest frame of one of the
two particles.

\section{The $\kappa$-exponential function}

We recall that the length of any four-vector is Lorentz
invariant. In particular for the four-momentum we have the
dispersion relation $p^{\mu} p_{\mu}=m^2c^2$ which can be
arranged as $(E/c-p)(E/c+p)=m^2c^2$ being $p=|{\bf p}|$. This
latter relationship after expressing the energy in terms of the
moment according to $E=\sqrt{m^2c^4 + p^2c^2}$ becomes
\begin{eqnarray}
\!\!\!\left(\!\sqrt{1\!+\!\left(\frac{p}{m c}\right)^2}-\frac{p}{m
c}\right)\!\left(\!\sqrt{1\!+\!\left(\frac{p}{m
c}\right)^2}+\frac{p}{m c}\right)=1 \ \ . \label{RIII6}
\end{eqnarray}
The latter equation, after introducing the dimensionless momentum
$q$, can be also written as
\begin{eqnarray}
\left(\!\sqrt{1\!+\!\kappa^2 q^2}-\kappa
q\right)^{1/\kappa}\!\left(\!\sqrt{1\!+\!\kappa^2 q^2}+\kappa
q\right)^{1/\kappa}\!\!=1  \ \ . \label{RIII8}
\end{eqnarray}
We remark that Eq. (\ref{RIII8}) follows directly from the
relativistic dispersion relation. Interestingly, in the classical
limit ${\kappa \rightarrow 0}$, Eq. (\ref{RIII8}) reduces to
$\exp(-q)\,\exp(q)=1$ while the dispersion relation becomes the
classical one $W=p^2/2m$ being $W$ the kinetic energy. In this
way we obtain a direct link between the dispersion relation of
free classical particles and the ordinary exponential function.
In the light of the above result, we reconsider Eq. (\ref{RIII8})
written in the form
\begin{eqnarray}
\exp_{_{\{{\scriptstyle \kappa}\}}}\!\left( -\,q \right)\,\,
\exp_{_{\{{\scriptstyle \kappa}\}}}\!\left( q \right) =1  \ \ ,
\label{RIII10}
\end{eqnarray}
with
\begin{eqnarray}
\exp_{_{\{{\scriptstyle \kappa}\}}}(q)=\left(\sqrt{1+\kappa^2 q^2
}+ \kappa q \right)^{1/\kappa} \ \ . \label{RI3}
\end{eqnarray}
We feel that the function $\exp_{_{\{{\scriptstyle
\kappa}\}}}\!\left(q \right)$, reproducing the ordinary
exponential function $\exp(q)=\exp_{_{\{{\scriptstyle
0}\}}}\!\left(q \right)$ in the classical limit $\kappa
\rightarrow 0$, represents a one parameter relativistic
generalization of the ordinary exponential.

Taking into account that the ordinary exponential results to be
eigenfunction of the ordinary derivative, namely
$(d/dq)\exp(q)=\exp(q)$, the question to determine the
eigenfunction of the $\kappa$-derivative, naturally arises. After
some simple calculation one obtains
\begin{eqnarray}
\frac{d}{d_{{\scriptscriptstyle\{}\kappa{\scriptscriptstyle\}}}q}
\,\exp_{_{\{{\scriptstyle \kappa}\}}}\!\!\left(q
\right)=\exp_{_{\{{\scriptstyle \kappa}\}}}\!\!\left(q \right) \ \
, \label{RIV13a}
\end{eqnarray}
so that the $\kappa$-exponential emerges as eigenfunction of the
$\kappa$-derivative
$d/d_{{\scriptscriptstyle\{}\kappa{\scriptscriptstyle\}}}q$.

The $\kappa$-exponential has the following important property
\begin{eqnarray}
\exp_{_{\{{\scriptstyle \kappa}\}}}\!\!\left(q_1
\right)\exp_{_{\{{\scriptstyle \kappa}\}}}\!\!\left(q_2
\right)=\exp_{_{\{{\scriptstyle \kappa}\}}}\!\!\left( q_1
\stackrel{\kappa}{\oplus} q_2\right) \ \ , \label{RIV13b}
\end{eqnarray}
being $q_1\stackrel{\kappa}{\oplus}q_2$ the additivity law of the
dimensionless relativistic momenta.

We note that Eq. (\ref{RIV13a}) admits as unique solution the
$\kappa$-exponential and then can be viewed as a differential
equation defining it. Analogously Eq. (\ref{RIV13b}) is a
functional equation univocally defining  the $\kappa$-exponential
function. In conclusion we have discussed three different ways to
introduce the $\kappa$-exponential. The first way is the more
physical one employing the relativistic dispersion relation. The
second and third ways are more mathematical and employ the
differential equation (\ref{RIV13a}) and the functional equation
(\ref{RIV13b}) respectively.

The mathematical properties of the $\kappa$-exponential can be
found in refs. \cite{PhA01,PRE02,PRE05}. One important property
of this function is its power low asymptotic behaviour when its
argument tends to $\pm \infty$.

\section{The relativistic entropy}

The Boltzmann-Gibbs-Shannon entropy of a classical particle
system is defined as the mean value of the minus logarithm of the
distribution function, namely $S(f\,)=-\langle
\,\ln(f)\rangle=-\int d^3p \,\,\,f\,\ln(f)$. In order to define
the entropy of a relativistic particle system, we employ the same
definition with the only difference that the ordinary logarithm
now is replaced by the $\kappa$-logarithm, defined as the inverse
function of the $\kappa$-exponential, namely
\begin{equation}
\ln_{_{\{{\scriptstyle \kappa}\}}} (f)=
\frac{f^{\kappa}-f^{-\kappa}}{2\kappa} \ \ .   \label{RI4}
\end{equation}
The $\kappa$-logarithm is a one parameter continuous deformation
of the ordinary logarithm which is recovered in the limit $\kappa
\rightarrow 0$. The reason of the modification of the classical
entropy is due to the fact that the $\kappa$-exponential and then
the $\kappa$-logarithm naturally emerge  within the special
relativity at the place of the ordinary exponential and logarithm
appearing in classical statistical mechanics.

The new entropy is given by
\begin{equation}
S(f\,)=-\langle \,\ln_{_{\{{\scriptstyle
\kappa}\}}}\!(f)\rangle=-\int d^3p
\,\,\,f\,\ln_{_{\{{\scriptstyle \kappa}\}}}\!(f) \ ,\label{RI5}
\end{equation}
and after maximization under the proper relativistic constrains,
yields the distribution
\begin{equation}
f= \alpha\exp_{_{\{{\scriptstyle
\kappa}\}}}\!\!\bigg(-\frac{\left(p^{\nu}+e A^{\nu}\!/c
\right)\,U_{\nu}-mc^2-\mu }{\lambda\,k_{_{B}}T}\bigg) \ \ ,
\label{RVIII6}
\end{equation}
with $\alpha=\left[(1-\kappa)/(1+\kappa)\right]^{1/2\kappa}$ and
$\lambda=\sqrt{1-\kappa^2}$. Being  both these parameters real,
$|\kappa|<1$ follows. The above distribution results to be quite
different with respect to the relativistic Maxwell-Boltzmann
distribution where the $\kappa$-exponential is replaced by the
ordinary exponential. The distribution (\ref{RVIII6}), in the
global rest frame where the hydrodynamic four vector velocity is
$U^{\nu}=(c,0,0,0)$ and in absence of external forces
($A^{\nu}=0$), simplifies to
\begin{equation}
f=\alpha \exp_{_{\{{\scriptstyle \kappa}\}}}\!\!\left(-\,
\frac{W-\mu}{\lambda\,k_{_{B}}T}\right) \ \ , \label{RVIII7}
\end{equation}
being $W$ the relativistic kinetic energy. The latter distribution
presents, for $W\rightarrow \infty$, power law tails, namely $f
\propto W^{-1/\kappa}$, in contrast to the Maxwell-Gibbs
distribution which decays exponentially. Moreover the
distribution (\ref{RVIII7}) fits very well the experimental data
of the cosmic ray spectrum \cite{PRE02}.

It is already known in the literature that the entropy (\ref{RI5})
has the following properties: i) it is non-negative and achieves
its maximum value at equiprobability, $f(p)=1/\Omega$ for $\forall
p$; and this value is $S=\ln_{_{\{{\scriptstyle
\kappa}\}}}\!\!\Omega$; ii) it is concave so that the system is
stable in thermodynamic equilibrium; iii) it satisfies the Lesche
stability condition and then it is physically meaningful; iv) it
generates a selfconsistent and coherent statistical mechanics.
Furthermore it results that: v) this statistical mechanics can be
obtained as stationary case of a kinetics governed by a
generalized relativistic Boltzmann equation; vi) this generalized
Boltzmann kinetics obeys the H-theorem which represents the
second law of thermodynamics.

\end{document}